\documentclass[aps,prd,nofootinbib,showpacs,floatfix,superscriptaddress,preprintnumbers]{revtex4}
\usepackage{amsmath}
\usepackage{amssymb}
\usepackage{epsfig}
\usepackage{graphicx}
\usepackage[center,footnotesize,hang]{subfigure}
\usepackage{pstricks}
\usepackage{pst-coil}
\DeclareMathAlphabet{\mathsc}{OT1}{cmr}{m}{sc}

\newcommand {\ignore}[1]{}

\def\10{$SO(10)$}
\def\21{SU(2) $\otimes$ U(1) }

\def\422{$SU(4) \otimes SU(2) \otimes SU(2)$}
\def\321{SU(3) $\otimes$ SU(2) $\otimes$ U(1)}
\def\gsim{\raise0.3ex\hbox{$\;>$\kern-0.75em\raise-1.1ex\hbox{$\sim\;$}}}
\def\lsim{\raise0.3ex\hbox{$\;<$\kern-0.75em\raise-1.1ex\hbox{$\sim\;$}}}

\def\lsim{\raise0.3ex\hbox{$\;<$\kern-0.75em\raise-1.1ex\hbox{$\sim\;$}}}
\def\gsim{\raise0.3ex\hbox{$\;>$\kern-0.75em\raise-1.1ex\hbox{$\sim\;$}}}

\def \znbb {0\nu\beta\beta}

\newcommand{\AddrAHEP}{%
  AHEP Group, Institut de F\'{\i}sica Corpuscular --
  C.S.I.C./Universitat de Val{\`e}ncia \\
  Edificio Institutos de Paterna, Apt 22085, E--46071 Valencia, Spain}

 \newcommand{\ba}{\begin{array}}
\newcommand{\ea}{\end{array}}
\relax

\def\321{$SU(3)\times SU(2)\times U(1)$}

\begin{document}

\preprint{IFIC/11-46}\preprint{RM3-TH/11-12}

\renewcommand{\Huge}{\Large}
\renewcommand{\LARGE}{\Large}
\renewcommand{\Large}{\large}
\def \znbb {$0\nu\beta\beta$ }
\def \nbb {$\beta\beta_{0\nu}$ }
\title{Fermion masses and mixing with tri-bimaximal in $SO(10)$ with type-I seesaw}  \date{\today}
\author{G. Blankenburg} \email{blankenburg@fis.uniroma3.it } \affiliation{Dipartimento di Fisica `E.~Amaldi', Universit\`a di Roma Tre
\\
INFN, Sezione di Roma Tre, I-00146 Rome, Italy}
\author{S. Morisi} \email{morisi@ific.uv.es} \affiliation{\AddrAHEP}
\date{\today}

\begin{abstract}
We study a class of models for tri-bimaximal neutrino mixing in $SO(10)$ grand
unified SUSY framework. Neutrino masses arise from both type-I and type-II 
seesaw mechanisms. We use dimension five operators in order to not spoil tri-bimaximal mixing
by means of type-I contribution in the neutrino sector.
We show that it is possible to fit all fermion masses and mixings including also the recent T2K result
as deviation from the tri-bimaximal.

\end{abstract}


\maketitle

\section{Introduction}
Neutrino mixing leads to large atmospheric angle (maximal), large solar angle (trimaximal)
and small reactor angle. In particular recently T2K collaboration has given indication of non zero
reactor angle \cite{T2K}. After such a result, the global fits of neutrino parameters give non zero
reactor angle at $3\sigma$:
\begin{equation}
\sin^2\theta_{13} =0.013^{+0.022}_{-0.012}\quad \mbox{\cite{Schwetz:2011zk}},\qquad
\sin^2\theta_{13} =0.025^{+0.025}_{-0.020}\quad \mbox{\cite{Fogli:2011qn}}.
\end{equation} 
This interesting result seems in contradiction with tri-bimaximal (TBM) mixing ansatz \cite{Harrison:2002er} that 
predicts zero reactor angle. However TBM solar and atmospheric mixing angles can 
be used as first approximation. Deviation from zero
reactor angle can arises in grand unified theory (GUT) like $SU(5)$ and $SO(10)$ from the
charged sector, see for instance\,\cite{Antusch:2011qg}. While neutrino mass matrix is
diagonalized from TBM unitary matrix, charged leptons are not diagonal giving deviation
to the TBM.
In this paper we consider such a possibility in the framework of a supersymmetric (SUSY) $SO(10)$ model. In this scenario charged leptons and CKM mixings are strongly related, we therefore consider the TBM as a good starting point to be corrected in general by small (CKM-like) deviations.
In Ref.\cite{mor} has been shown that in a renormalizable $SO(10)$ model this is  not possible
in case of type-I seesaw. 
Such a  difficulty arises from the fact that  up quark and Dirac neutrino
Yukawa couplings are strongly related in renormalizable $SO(10)$ models.  
Some interesting attempts to obtain TBM with a flavour symmetry are developed in Ref.\cite{Dutta:2009bj} 
and Ref.\cite{king}, assuming type-II seesaw to be dominant 
\footnote{For an incomplete list of papers with TBM in GUT see \cite{guttbm}, for $SO(10)$ models with discrete flavour symmetry and no TBM mixing see \cite{Dermisek:1999vy} and for a collection of general $SO(10)$ models see at the References in Ref. \cite{albr}}.
In Ref.\cite{bla}, in the context of SUSY renormalizable $SO(10)$ with type-II seesaw dominance, a fit of all the fermion 
masses and mixing  has been done (see also \cite{josh}). The superpotential considered is of the form
\begin{equation}\label{w0}
w= h 16\,16\,10+f 16\,16\,\overline{126}+h' 16\,16\,120,
\end{equation}
where $h$ is a symmetric matrix, $h'$ is antisymmetric and $f$ has the TBM structure, namely
\begin{equation}\label{TBMt}
f=\left(
\begin{array}{ccc}
f_2&f_1&f_1\\
f_1&f_2+f_0&f_2-f_0\\
f_1&f_2-f_0&f_2+f_0\\
\end{array}
\right),
\end{equation}
where $m_{\nu 1}=f_2-f_1$, $m_{\nu 2}=f_2+2f_1$ and $m_{\nu 3}=f_2-f_1+2 f_0$. 
It is well know that a mass matrix with the above structure is diagonalized by TBM mixing
matrix, see for instance\,\cite{Altarelli:2010fk}. No assumptions have been made taking $f$ to be TBM because we can always go to this basis by rotating the 16 of fermions \cite{bla}.
The matrices $h,h'$ are assumed to be hermitian that can correspond to an underlying parity \cite{moh}. 

Another possibility to reproduce TBM mixing in the framework of $SO(10)$ GUT models 
is to use non-renormalizable operators containing a scalar field transforming as a $45_H$
of $SO(10)$ \cite{deMedeirosVarzielas:2006fc,Morisi:2007ft,mor}.
This field allows to distinguish up quarks from neutrinos permitting TBM mixing also in the case where 
neutrino masses arise from  type-I seesaw mechanism. In particular in Ref.\cite{mor} for this
purpose the dimension five operator $16\,16\,120_H\,45_H$ has been used.  
This operator yields a contribution to the up-quark mass matrix 
and not to the Dirac neutrino one allowing to distinguish the up-quark from Dirac neutrino sectors. In this way it is possible to obtain both Dirac and Majorana neutrino masses TBM and hierarchical structure in charged fermions sector. 


A full fit of quark and lepton masses and mixing in models with TBM mixing from type-I
seesaw in $SO(10)$ is still missing. 
In this paper we consider such a problem. 
We link the idea of distinguish up-quark and Dirac neutrino by means of $16\,16\,120_H\,45_H$ operator
with the result of Ref.\,\cite{bla} where has been shown that from the superpotential (\ref{w0}) it is possible
to fit all the data having TBM mixing in the neutrino sector.
In this paper we will translate the superpotential\,(\ref{w0}) in the language of dimension five operators.

\vskip5.mm
In the next section we will review some of the $SO(10)$ dimension five operator that will be useful
to construct an $SO(10)$ model giving TBM mixing with type-I as well as type-II seesaw
following the indication given in the superpotential\,(\ref{w0}) of Ref.\,\cite{bla}.
In section\,\ref{3} we give some examples of models and the corresponding fits, in section\,\ref{4}
we discuss the possibility to obtain a renormalizable model, then in section\,\ref{5} 
we give our conclusions.

\section{dimension five effective operators}\label{2}
In this section we report the main ingredients that will be useful to construct our model in 
the next section. It contains some of the result of table VIII of Ref.\cite{mor} that we report in appendix\,A for 
the useful of the reader.

As discussed in the introduction, one possibility to reproduce TBM mixing in $SO(10)$ in the case of
type-I seesaw is by means of the dimension five operator $16\,16\,120_H\,45_H$ that allows to 
distinguish between up-quark  and Dirac neutrino sectors. 
In this section we remark the feature of some dimension five operators that we will 
use in the next section.

In general an $SO(10)$ dimension five operator can be written  as $16\,16\,\phi_a\, \phi_b$ 
where $\phi_{a,b}$ are scalar fields $\phi_{a,b}=1_H,\,16_H,\, \overline{16}_H,\, 
45_H,...$ and so on. 
For simplicity we assume that $SO(10)$ is broken through $SU(5)$ and we describe
the contribution of the dimension five operators to the fermion mass matrices in the
$SU(5)$ language. 
When one of the components of $\phi_a$ and $\phi_b$ take vev $a_i$ and $b_i$ 
respectively (where $i$ is the $SU(5)$ index of the component), one generates contributions
to the quark and lepton masses. Note that $a_i$  and $b_i$ can be possibly equal if 
$\phi_a=\phi_b$.

\vskip4.mm
The dimension five operators that will be used are: 
\begin{itemize}

\item $16\,16\,\overline{16}_H\,\overline{16}_H$ 
\vskip2.mm
This operator can be obtained, for example, integrating out a $SO(10)$ singlet $1_\chi$ or a 45-plet $45_\chi$ of heavy messenger fermions:
\begin{equation}\label{16b16b}
f\,(16\overline{16}_H)_1(16\overline{16}_H)_1,\qquad f \,(16\overline{16}_H)_{45}(16\overline{16}_H)_{45}.
\end{equation}
>From the table in appendix A it is possible to see that the first operator contributes to 
$Y_\nu$, $M_R$ and $M_L$ while the second one contributes to $Y_\nu$, $M_R$, $M_L$ and $Y_u$.

We assume $f$ to have the  TBM form\,(\ref{TBMt}), in this way the light neutrino mass is TBM with type-I or 
type-II seesaw, see also eq.\,(\ref{mnu}) below. Note that $f$ TBM is general because we can always go in this basis by rotating the 16 \cite{bla}. 

\item $16\,16\,120_H\,45_H$
\vskip2.mm
This operator can be obtained by integrating out a couple $16_\chi-\overline{16}_\chi$
\begin{equation}\label{12045}
h\,(16\,120_H)_{16}(16\,45_H)_{\overline{16}}.
\end{equation}
It can yield a contribution to the up-quark mass matrix (and to the down-quark and charged lepton mass matrices) 
and not to the Dirac neutrino one, allowing to distinguish the up-quark from Dirac neutrino sectors. 

This can be described naively in the $SU(5)$ language as follows. 
The up-type Higgs doublet in a $45_H$ of $SU(5)$ (contained into
the $120_H$) couples antisymmetrically to the two matter multiplets $10$ that give the up-quark mass $M_u$
while it does not contribute to the Dirac neutrinos, when contracted as in eq. (\ref{12045}). 
This is a well know feature of $SO(10)$ where the operator $16 \,16\, 120_H$ contributes to $M_u\propto \langle 45_{SU(5)} \rangle$ and 
$M_\nu\propto \langle 5_{SU(5)} \rangle$ with antisymmetric Yukawa. But with the dimension five operator the resulting mass matrix is
not antisymmetric and so can give the hierarchical structure of the quark sector. In fact with the insertion of a $24_H$ scalar multiplet (contained in the $45_H$ of $SO(10)$
) that takes vev in the hypercharge direction, results that $M_u$ is not antisymmentric (but a generic matrix) since the Clebsch-Gordan coefficients for
the isospin doublet $Q$ and the isosinglet $u^c$ are different with a relative factor $(-4)$ that arise
from the hypercharge. 

We can describe the property of the $16\,16\,120_H\,45_H$ operator in more detail as follow. 
The $45_H$ can take vev in its singlet $1_{SU(5)}$ component 
called $X$-direction\footnote{This is the extra $U(1)$ contained in $SO(10)\supset SU(5)\times U_X(1)$.}  
or along the adjoint $24_{SU(5)}$ component, that is the hypercharge $Y$-direction  
(see for instance \cite{Anderson:1993fe}). 
We indicate their vevs as
\begin{equation}\label{v124}
b_1=\langle 1_{SU(5)} \rangle,\qquad b_{24}=\langle 24_{SU(5)} \rangle.
\end{equation}
Equivalently, the $45_H$ can take vev along the isospin direction or the $B-L$ direction
and their corresponding vev are denoted as $b_3$ and $b_{15}$ respectively
and are given by
\begin{equation}\label{v315}
b_1=\frac 15 (b_3+3b_{15}) ~,~~~~~ b_{24}=\frac 15 (-b_3+2b_{15}) ~.
\end{equation}
The $SU(5)$ components of the $120_H$ of $SO(10)$ that contain $SU(2)$ doublet 
(giving rise to the Dirac masses terms for the fermions)
are the $45_{SU(5)}$, $\overline{45}_{SU(5)}$, $\overline{5}_{SU(5)}$ and $5_{SU(5)}$ 
representations. We denote their vevs as 
\begin{equation}\label{ai}
a_5=\langle 5_{SU(5)} \rangle,\qquad 
a_{\bar{5}}=\langle \overline{5}_{SU(5)} \rangle,\qquad 
a_{45}=\langle 45_{SU(5)} \rangle,\qquad 
a_{\bar{45}}=\langle \overline{45}_{SU(5)} \rangle. 
\end{equation}
>From the table in appendix A we have for instance that
\begin{eqnarray}
M_u &=&h\, a_{45}(b_1-4b_{24})-h^T\, a_{45}(b_1+b_{24}),\\
M_\nu&=& 5\, h\, a_{5}b_1 - h^T\, a_{5}(-3b_1-3b_{24}),\\
M_d&=&h (a_{\overline{5}}+a_{\overline{45}})(-3b_1+2b_{24})
-h^T (a_{\overline{5}}+a_{\overline{45}})(b_1+b_{24}) ,\\
M_e^T&=&h (a_{\overline{5}}-3 a_{\overline{45}})(-3b_1-3b_{24})
-h^T (a_{\overline{5}}- 3 a_{\overline{45}})(b_1+6b_{24}) .
\end{eqnarray} 
Then if $a_5=0$ this operator contributes to $Y_u$, $Y_d$, $Y_e$ and not to $Y_\nu$.
So $h$ can be in a hierarchical form, with $(3,3)$ element dominant, as required by charged fermion phenomenology,
without changing the TBM result. 
Because of the $45_H$, the mass matrix that results from such an operator is not antisymmetric but general. 
Note that with a $45_H$ in B-L direction the resulting mass matrix  is symmetric, in fact 
using eq.\,(\ref{v315}) we have
\begin{eqnarray}
M_u &=&h\, a_{45}(b_3-b_{15})-h^T\, a_{45}(b_{15}),\\
M_\nu&=&  h\, a_{5}(b_3+3 b_{15}) + h^T\, a_{5}(3b_{15}),\\
M_d&=&h(a_{\overline{5}}+a_{\overline{45}})(-b_3-b_{15})-h^T (a_{\overline{5}}+a_{\overline{45}})b_{15} ,\\
M_e^T&=&h(a_{\overline{5}}-3a_{\overline{45}})(-3b_{15})-h^T (a_{\overline{5}}-3a_{\overline{45}})(3b_{15}-b_3) .
\end{eqnarray} 
and stetting $b_3=0$ then all the mass matrices are symmetric.

\item $16\,16\,10_H\,45_H$
\vskip2.mm
This operator can be obtained by contracting with a couple $16_\chi-\overline{16}_\chi$:
\begin{equation}\label{1045}
h^\prime(16\,10_H)_{16}(16\,45_H)_{\overline{16}}\,,
\end{equation}
We denote the vev of the $SU(5)$ components of the $10_H$, containing all the possible $SU(2)$ doublets, as 
\begin{equation}
a_5=\langle 5_{SU(5)} \rangle,\qquad 
a_{\bar{5}}=\langle \bar{5}_{SU(5)} \rangle.
\end{equation}
If $a_5=0$ this operator contributes only  in $Y_e$ and $Y_d$ as usual in $SU(5)$. In fact from Appendix A we have
\begin{eqnarray}
M_u &=&h^\prime a_5 (b_1 - 4 b_{24}) + h^{\prime T} a_5 ( b_1 + b_{24}) ,\\
M_\nu&=&5 h^\prime a_5 b_1 + h^{\prime T} a_5 (-3 b_1 - 3 b_{24}) ,\\
M_d&=&h^\prime a_{\overline 5} (-3b_1 + 2 b_{24})
+ h^{\prime T} a_{\overline 5} (b_1 + b_{24}) ,\\
M_e^T&=&h^\prime a_{\overline 5} (-3b_1 - 3 b_{24})
+ h^{\prime T} a_{\overline 5} (b_1 + 6 b_{24}) .
\end{eqnarray} 
Again, because of the $45_H$, the resulting mass matrix is not symmetric but a generic matrix and it can 
contribute to the down-quark and lepton masses. Note that with a $45_H$ in B-L direction the resulting mass 
matrix is antisymmetric
in fact
\begin{eqnarray}
M_u&=&h^\prime a_5 (b_3 - b_{15}) + h^{\prime T} a_5  b_{15} ,\\
M_\nu&=&5 h^\prime a_5 (b_3 + 3b_{15}) - h^{\prime T} a_5 3 b_{15} ,\\
M_d   &=& h' a_{\overline 5} (-3b_3 -  b_{15}) + h'^T a_{\overline 5}  b_{15}, \\
M_e^T &=& h' a_{\overline 5} (-3 b_{15}) + h'^T a_{\overline 5} (-b_3 +3 b_{15}). 
\end{eqnarray}
and putting $b_3=0$ the mass matrices are clearly antisymmetric.

\item Adding an $SO(10)$ scalar singlet $1_H$ we can consider also the dimension five operators
\begin{equation}
16\,16\,\overline{126}_H\,1_H,\qquad 16\,16\,10_H\,1_H,\qquad 16\,16\,120_H\,1_H\,,
\end{equation} 
that behave in the same way as the renormalizable ones (see appendix A) introduced in eq.\,(\ref{w0}). 
Also these operators can be obtained integrating out a $16_\chi-\overline{16}_\chi$ of heavy messenger fermions
\begin{equation}\label{X1}
(16\,\overline{126}_H)_{16}(16\,1_H)_{\overline{16}},\qquad (16\,10_H)_{16}(16\,1_H)_{\overline{16}},\qquad 
(16\,120_H)_{16}(16\,1_H)_{\overline{16}}\,.
\end{equation}

\end{itemize}

We see that the key ingredient to obtain type-I seesaw and TBM mixing is that the up-type $SU(2)$ Higgs
doublets in the $5_{10}$ and $5_{120}$ do not have vevs and so that they are not in linear combination
of the light Higgs doublet. This can be a potentially problem since $\bar{5}_{10}$ takes a vev and it 
is mixed with the other light Higgs doublets. However the study of the complete scalar potential is beyond the scope
of this paper and will be studied elsewhere.  

\section{models for TBM and fit of fermion masses and CKM}\label{3}

In Ref.\,\cite{bla} has been studied a model for TBM mixing with dominant type-II seesaw mechanism given in eq.\,(\ref{w0}).
In this section we present some possible modifications of the model given in eq.\,(\ref{w0}). In the
models we will present below, TBM arises from both type-I and type-II seesaw mechanisms differently from 
Ref.\cite{bla,Dutta:2009bj} where dominant type-II seesaw mechanism has been assumed for neutrino masses. 
We remark that the main problem with type-I seesaw is that the tree-level operator 
$16\, 16\, 10$ gives equal contribution to the up-quark and Dirac neutrino mass matrix. 
But in order to fit quark masses and mixings with TBM neutrino mixing, the structure of the two mass matrices must 
be very different, namely the up quark mass matrix must be hierarchical while the Dirac neutrino mass 
matrix must be of TBM-type as in eq.\,(\ref{TBMt}) or the identity. So we need to disentangle the two sectors, leaving Dirac and Majorana neutrino masses of TBM-type and Dirac charged fermions masses hierachical and almost diagonal.  
>From the previous section it is clear that one possibility is to replace the operator $16 \,16 \,10$ of eq.\,(\ref{w0}) 
with the operator $16\, 16\, 45\, 120$.

In the following we will assume an underlying parity, like in \cite{moh}, making all the mass matrices hermitian
and so reducing the number of free parameters.
Another way to reduce the sometimes high number of parameters is to assume that the 45 get vev in the B-L direction. 
In this case the fermion mass matrices are symmetric or antisymmetric and not arbitrary.

\vskip4.mm
Examples of models with TBM neutrino mixing are listed below. The details of the fit are given in appendix B and C.
We fit all charged fermion masses, the two neutrino mass square differences, leptons and quarks mixings,
and the CKM phase for a total of 18 observables.
For the operators $16\,16\,120_H\,45_H$ and $16\,16\,10_H\,45_H$ we always take zero vev for the component $5_{SU(5)}$ of $120_H$ and $10_H$, as described in the previous section ($a_5=0$).

\begin{itemize}

\item {\it case A}: $w=f\, 16\,16 \,\overline{16}_H\, \overline{16}_H + h\, 16 \,16 \,45_H\, 120_H + h'\, 16\, 16\, 45_H\, 10_H  $ 
\vskip2.mm
where the $45_H$ takes vev in a general direction, that is $b_{1}$ and $b_{24}$ 
(see eq.\,(\ref{v124})), are both different from zero. The mass matrices are :
\begin{eqnarray}
M_u &=& h+f,\\
M_d &=& r_1[h(2\frac{b_1}{b_{24}}-3)+h^T(2\frac{b_1}{b_{24}}+2)+h^\prime],\\
M_e &=& r_1\{c_e[h(2\frac{b_1}{b_{24}}+7)+h^T(2\frac{b_1}{b_{24}}+2)]+h^\prime(-5\frac{b_{24}}{4b_1-b_{24}})+ h^{\prime T}
(4\frac{b_1+b_{24}}{4b_1-b{24}})\},\\
M_{\nu^D} &=& \frac{1}{2}f,\\
M_R &=& r_R f,\\
M_L &=& r_L f,
\end{eqnarray}
where $h$ and $h^\prime$ are generic matrices, $r_i$, $c_e$ and $b_i$ are combinations of vevs (see appendix A).

Results:
\begin{equation}
\chi^2 = 0.0050, 
\end{equation}
with 26 parameters. 

We note that with the $45_H$ taking vev in B-L direction the number of parameters is considerably reduced but a good fit can not be performed.

%
%
%
%
%
%

\item {\it case B}: $w=f\, 16\,16 \,\overline{126}_H\, 1_H + h\, 16 \,16 \,45_H\, 120_H + h'\, 16\, 16\, 45_H\, 10_H  $ 
\vskip2.mm
where the $45_H$ takes vev in B-L direction.
The mass matrices are:
\begin{eqnarray}
M_u &=& h^S+r_2f,\\
M_d &=& r_1(h^S+h^A+f),\\
M_e &=& r_1(c_eh^S-3h^A-3f),\\
M_{\nu^D} &=& -3r_2f,\\
M_R &=& r_R f,\\
M_L &=& r_L f.
\end{eqnarray}

Results:
\begin{equation}
\chi^2 = 5.6,
\end{equation}
with 16 parameters (2 d.o.f).

%
%
%

\item {\it case C}: $w=f\, 16\,16 \,\overline{126_H}\, 1_H + h\, 16 \,16 \,45_H\, 120_H + h'\, 16\, 16\, 120'_H\,1_H  $ 
\vskip2.mm
where the $45_H$ takes vev in B-L direction.
The mass matrices are:
\begin{eqnarray}
M_u &=& h^S+r_3h^A+r_2f,\\
M_d &=& r_1(h^S+h^A+f),\\
M_e &=& r_1(c_e^Sh^S+c_e^Ah^A-3f),\\
M_{\nu^D} &=& -3r_2f,\\
M_R &=& r_R f,\\
M_L &=& r_L f.
\end{eqnarray}

Results:
\begin{equation}
\chi^2 = 0.0015,
\end{equation}
with 18 parameters.

\end{itemize}

The last case reproduces basically the same structure of the renormalizable case (eq. (\ref{w0})) with type-II seesaw dominance studied for example in ref. \cite{bla}, with just one more parameter $c_e^S$. We note that the analysis performed in \cite{bla} is based on a previous set of data (before the T2K and MINOS recent results). For this reason we show also an updated fit for that case, that can be used for comparison:
\begin{equation}
\chi^2 = 0.14, \qquad d_{FT} = 461863
\end{equation}
with 17 parameters (1 d.o.f), where $d_{FT}$ is a parameter introduced in \cite{bla}.
We note that the goodness of the fit is substantially unchanged compared with the old analysis, showing that in this class of models it is possible to obtain the desired (very small before T2K or more sizeable now) corrections to zero $\theta_{13}$ from the charged lepton sector, taking into account an appreciable amount of finetuning. 
In fact the neutrino mass matrix is of TBM-type and  it is diagonalized by TBM mixing matrix. 
The charged fermion mass matrices have hierarchical structure. 
Assuming all the parameters to be real, the charged lepton 
mass matrix is diagonalized by a rotation matrix $O^l$ characterized by three angles 
$\theta_{13}^l$, $\theta_{12}^l$ and $\theta_{23}^l$. The lepton mixing matrix $V_{lep}$ is given
by the product $V_{lep}={O^l}^\dagger \cdot V_{TBM}$ so we have
\begin{eqnarray}
{(V_{lep})}_{13}&=&\frac{1}{\sqrt{2}} (s_{13}-c_{13} s_{12}),\\
{(V_{lep})}_{12}&=&\frac{1}{\sqrt{3}} (c_{12} c_{13}+ s_{12} c_{13}+s_{13}),\\
{(V_{lep})}_{23}&=&\frac{1}{\sqrt{2}} (-c_{12} c_{23}+ s_{23}(c_{13} +s_{12}s_{13})),
\end{eqnarray}
where $s_{ij}=\sin\theta_{ij}^l$ and $c_{ij}=\cos\theta_{ij}^l$.
We can have a large value for the reactor angle in agreement with the result of the T2K collaboration,
and at the same time  ${(V_{lep})}_{12} \approx 1/\sqrt{3}$ and 
${(V_{lep})}_{23} \approx 1/\sqrt{2}$ fine-tuning the mixing angles $\theta_{ij}^l$. 

\vskip4.mm
We observe that from type-I and type-II seesaw mechanisms we have for all the cases presented above
\begin{equation}\label{mnu}
m_\nu=M_L-M_{\nu^D}\frac{1}{M_R}M_{\nu^D}^T=(r_L-\frac{1}{r_R})\,f=r_\nu\,f\, ,
\end{equation}
where we have used the fact that $f=f^T$. Note that only a combination of the $r_{L,R}$ 
parameters enters in the neutrino sector. So counting the number of free parameters, $r_L$ and $r_R$ are equivalent to one free parameter instead
of two.


\section{Renormalizable theory}\label{4}

The dimension five operators assumed in the previous section can be obtained from a renormalizable theory 
integrating out heavy messengers fields.
In general the operator $16 \,16 \,\phi_a\,\phi_b$ can be obtained from 
\begin{equation}
w= 16 \,16_\chi \phi_a + 16 \,\overline{16}_\chi \phi_b + M_\chi16_\chi\,\overline{16}_\chi 
\end{equation}
where $\overline{\chi}-\chi$ is a couple of sets of fermion messengers\footnote{
If only one messenger field is assumed the effective Yukawa mass matrix is rank one.} and it gives rise to  the operator 
 $16 \,16 \,\phi_a\,\phi_b$ at a scale $E\ll M_{\chi}$.

Moreover it is easy to take a symmetry forbidding the direct tree level operators $16\,16\,\phi_a$, for $\phi_a=10_H,\,120_H,\,\overline{126}_H$. For example we can take a $Z_2$ symmetry acting as 
\begin{equation}
(16, \phi_{a,b}) \rightarrow (16, -\phi_{a,b}),\qquad (\chi,\overline{\chi}) \rightarrow -(\chi,\overline{\chi}).
\end{equation}

\vskip4.mm
Below we report explicit examples of renormalizable models from which the effective dimension five
superpotentials assumed in the previous section can be obtained:
\begin{itemize}
\item case {\it A}\\
The matter and scalar fields content of a possible renormalizable model that can give the 
effective superpotential of the case {\it A} is given by:
\begin{center}
\begin{tabular}{|c|c|cccc||cccc|}
\hline
 &$16$&$10_H$&$\overline{16}_H$&$45_H$&$120_H$&$1_\chi$&$16_\chi$&$\overline{16}_\chi$& $45_\chi$\\
$Z_2$ & $+$  & $-$  & $-$  & $-$  & $-$  & $-$  & $-$  & $-$ & $-$ \\
\hline
\end{tabular}
\end{center}
then the renormalizable superpotential is 
\begin{equation}
w=16 \,1_\chi \,\overline{16}_H +16 \,45_\chi \,\overline{16}_H+ 16 \,16_\chi \, 120_H+ 16\,\overline{16}_\chi 45_H +
16 \,16_\chi \, 10_H\,+M\,16_\chi\overline{16}_\chi +M_1 1_\chi1_\chi+M_{45} 45_\chi45_\chi.
\end{equation}


\item case {\it B}\\
The matter and scalar field content of the model is   
\begin{center}
\begin{tabular}{|c|c|ccccc||cccc|}
\hline
 &$16$&$1_H$&$10_H$&$45_H$&$120_H$& $\overline{126}_H$ &$16_{\chi 1}$&$\overline{16}_{\chi1}$ &$16_{\chi2}$&
$\overline{16}_{\chi2}$\\
$Z_2$ & $+$  & $-$  & $-$  & $-$  & $-$  & $-$  & $-$  & $-$ & $-$ & $-$ \\
$Z'_2$ & $+$  & $+$  & $-$  & $-$  & $-$  & $+$  & $+$  & $+$ & $-$ & $-$ \\
\hline
\end{tabular}
\end{center}
and the superpotential is given by 
\begin{equation}
w=16\,16_{\chi 1}\,\overline{126}_H + 16 \,\overline{16}_{\chi 1} 1_H+16\,16_{\chi 2}\,10_H+16\,16_{\chi 2}\,120_H
+ 16 \,\overline{16}_{\chi 2} 45_H+M_1\, 16_{\chi 1}\overline{16}_{\chi 1}+M_2\, 16_{\chi 2}\overline{16}_{\chi 2}
\end{equation}

\item case {\it C}\\
The matter and scalar field content of the model is   
\begin{center}
\begin{tabular}{|c|c|ccccc||cccc|}
\hline
 &$16$&$1_H$&$120_H$&$45_H$&$120'_H$& $\overline{126}_H$ &$16_{\chi 1}$&$\overline{16}_{\chi1}$ &$16_{\chi2}$&
$\overline{16}_{\chi2}$\\
$Z_2$ & $+$  & $-$  & $-$  & $-$  & $-$  & $-$  & $-$  & $-$ & $-$ & $-$ \\
$Z'_2$ & $+$  & $+$  & $-$  & $-$  & $+$  & $+$  & $+$  & $+$ & $-$ & $-$ \\
\hline
\end{tabular}
\end{center}
and the superpotential is given by
\begin{equation}
w=16\,16_{\chi 1}\,\overline{126}_H + 16 \,\overline{16}_{\chi 1} 1_H+16\,16_{\chi 1}\,120'_H+16\,16_{\chi 2}\,120_H
+ 16 \,\overline{16}_{\chi 2} 45_H+M_1\, 16_{\chi 1}\overline{16}_{\chi 1}+M_2\, 16_{\chi 2}\overline{16}_{\chi 2}\,.
\end{equation}

\end{itemize}

\section{Conclusions}\label{5}

Neutrino mixing data are in well agreement with maximal atmospheric angle,
tri-maximal solar angle and may be with a non-zero and quite large (namely 
of order of the Cabibbo angle) reactor angle. TBM mixing gives zero reactor angle
however it can be a reasonable starting point. 
In fact in GUT framework large deviation of the $1-3$  angle can arise from the charged
sector. However a simple picture for TBM in $SO(10)$ is still missing.
In order to approach the problem recently has been studied models where 
light-neutrino mass matrix arises only from type-II seesaw mechanism.
In this paper we studied the possibility that both type-I and type-II seesaw mechanisms 
yield TBM neutrino mixing in a $SO(10)$ model. We have assumed that the superpotential contains only dimension 
five non-renormalizable operators. We studied three different possible scenarios for TBM neutrino mixing
In each case proposed we make the fits of all the fermion masses and mixing angle. 
One case corresponds to the model studied already in Ref.\,\cite{bla} for type-II seesaw dominance, while the other two are new.

We found in both cases a good fit of all the data including the recent T2K result. 
In particular for the first model we found an excellent fit ($\chi^2=0.005$) but with a high number (26) of parameters. We therefore can conclude that this case can be considered as a good starting point for a flavour theory that can reduce the number of the free parameters of the theory (for example introducing a flavour symmetry). Moreover in this case we did not need to introduce extra $SO(10)$ singlets.
For the second case we obtained $\chi^2=5.6$ but with only 16 free parameters and 2 degree of freedom, making this case the most predictive. For third case we found a very good fit $\chi^2=0.002$ with 18 free parameters. This case also can be considered as a good starting point for a complete flavour theory. Even if we needed to introduce one $SO(10)$ singlet we consider the last two cases as the most promising for the moment.

For the three cases proposed, we give possible renormalizable realizations
where we have introduced messenger fields and extra Abelian symmetries.

We remark that in this paper we focused on the flavour secotr and we do not make a full analysis of the model. In particular we leave to a future analysis the study of the Higgs potential and related issues such as the breaking pattern of $SO(10)$ to the SM, problems related to the doublet-triplet splitting (proton-decay) and the achieving of exact coupling unification considering the possible breaking steps and the related threshold corrections.

\section{Acknowledgments}

We thank Prof.\,G.Altarelli for the useful comments. This work was supported by the Spanish MICINN under grants
FPA2008-00319/FPA, FPA2011-22975 and MULTIDARK CAD2009-00064 (Con-solider-Ingenio
2010 Programme), by Prometeo/2009/091 (Generalitat Valenciana), by the
EU Network grant UNILHC PITN-GA-2009-237920.
G.B. thanks the AHEP group for partial support during his visit to Valencia for the FLASY workshop 
where this project has been started.

\newpage
\section*{Appendix A}

Here we report for convenience of the reader the table VIII of ref.\cite{mor}.
%
In general an $SO(10)$ dimension five operator can be written  as $16\,16\,\phi_a\, \phi_b$ 
where $\phi_{a,b}$ are scalar fields $\phi_{a,b}=1_H,\,16_H,\, \overline{16}_H,\, 
45_H,...$ and so on. 
For simplicity we assume that $SO(10)$ is broken through $SU(5)$ and we describe
the contribution of the dimension five operators to the fermion mass matrices in the
$SU(5)$ language. 
When one of the components of $\phi_a$ and $\phi_b$ take vev $a_i$ and $b_i$ 
respectively (where $i$ is the $SU(5)$ index of the component), one generates contributions
to the quark and lepton masses.\\ 
\begin{small}
\begin{center}
\begin{table}[hbtp]
\begin{tabular}{|c|c|c|}
\hline
case & $SO(10)$ operator & mass matrices  \\
\hline
\hline 
IV & $(16_M 16_H)_{10} (16_M 16_H)_{10}$ &
$\begin{array} {l}  
M_d=     K a_{\overline{5}} b_1 + K^T a_1 b_{\overline 5} \\
M_e^T= K a_{\overline{5}} b_1 + K^T a_1 b_{\overline 5} \\
\end{array}$\\
\hline
V & $(16_M \overline{16}_H)_1 (16_M \overline{16}_H)_1$ & 
$\begin{array} {l}  
M_\nu = K a_5 b_1 + K^T a_1 b_5 \\
M_L = K_s a_5 b_5 \\
M_R = K_s a_1 b_1 \\
\end{array}$\\
\hline 
VI & $(16_M \overline{16}_H)_{45} (16_M \overline{16}_H)_{45}$ &
$\begin{array}{l} 
M_u = 8 K_s (a_5 b_1 + a_1 b_5) \\
M_\nu = 3 (K a_5 b_1 + K^T a_1 b_5) \\
M_L = -5 K_s a_5 b_5 \\
M_R = -5 K_s a_1 b_1 \\
\end{array}$\\
\hline
VII & $(16_M 10_H)_{16} (16_M 45_H)_{\overline{16}}$ & 
$\begin{array}{l} 
M_u = K a_5 (b_1 - 4 b_{24}) + K^T a_5 ( b_1 + b_{24}) \\
M_\nu = 5 K a_5 b_1 + K^T a_5 (-3 b_1 - 3 b_{24})  \\
M_d = K a_{\overline 5} (-3b_1 + 2 b_{24})
+ K^T a_{\overline 5} (b_1 + b_{24}) \\
M_e^T = K a_{\overline 5} (-3b_1 - 3 b_{24})
+ K^T a_{\overline 5} (b_1 + 6 b_{24}) \\
\end{array}$\\
\hline
VIII & $(16_M 120_H)_{16} (16_M 45_H)_{\overline{16}}$ & 
$\begin{array} {l} 
M_u =K a_{45}(b_1-4b_{24})-K^T a_{45}(b_1+b_{24})\\
M_\nu= 5 K a_{5}b_1 - K^T a_{5}(-3b_1-3b_{24})\\
M_d =K (a_{\overline{5}}+a_{\overline{45}})(-3b_1+2b_{24})
-K^T (a_{\overline{5}}+a_{\overline{45}})(b_1+b_{24})\\
M_e^T=K (a_{\overline{5}}-3 a_{\overline{45}})(-3b_1-3b_{24})
-K^T (a_{\overline{5}}- 3 a_{\overline{45}})(b_1+6b_{24})\\
 \end{array} $ \\
\hline
IX & $(16_M 16_H)_{120} (16_M 16_H)_{120}$ & 
$\ba{l}
M_d = K (a_{\overline{5}} b_1 + 2 a_1 b_{\overline{5}} )
+ K^T (a_1 b_{\overline{5}} +2 a_{\overline{5}} b_1)\\
M_e^T = K (a_{\overline{5}} b_1 + 2 a_1 b_{\overline{5}} )
+ K^T (a_1 b_{\overline{5}} +2 a_{\overline{5}} b_1)\\
\ea$
\\ \hline
\end{tabular}
\caption{
The contributions to the mass matrices from $SO(10)$-invariant dim-5 operators,
from table VIII of ref.\cite{mor}. $K$ is an arbitrary matrix.}
\label{tab}
\end{table}
\end{center}
\end{small}

\vskip4.mm
Below we report the contributions to the mass matrices from $SO(10)$ invariant 
renormalizable Yukawa couplings. 
Different VEVs of the same $SO(10)$ Higgs multiplet carry a subscript indicating the $SU(5)$ component they belong to.
\begin{small}
\begin{center}
\begin{tabular}{|c|c|c|}
\hline 
case & $SO(10)$ operator & mass matrices \\
\hline \hline
I & $ 16_M 16_M 10_H$ & 
$\begin{array}{l}
M_u=M_{\nu}= Y_{10} v_5 \\
M_d=M_e = Y_{10} v_{\overline{5}} 
\end{array}$\\
\hline
II & $16_M 16_M 120_H$ & 
$\begin{array}{l}
M_u= Y_{120} v_{45} \\
M_{\nu} = Y_{120} v_5 \\
M_d = Y_{120} (v_{\overline{5}}+v_{\overline{45}}) \\
M^T_e = Y_{120} (v_{\overline{5}}-3v_{\overline{45}})
\end{array}  $\\
\hline
III & $16_M 16_M \overline{126}_H$ & 
$ \begin{array}{l}
M_u= Y_{\overline{126}}\, v_5 \\ 
M_{\nu} = -3 Y_{\overline{126}}\, v_5 \\  
M_d = Y_{\overline{126}}\, v_{\overline{45}} \\ 
M_e = -3 Y_{\overline{126}}\, v_{\overline{45}} \\ 
M_L = Y_{\overline{126}}\, v_{15} \\
M_R = Y_{\overline{126}}\, v_1
\end{array}$\\
\hline
\end{tabular}
\end{center}
\end{small}

\newpage

\section*{Appendix B}

In this section we show the fitting procedure used in our analysis.
For charged fermions and CKM mixings the fit are performed on the set of data evolved at the GUT scale showed in Tab. \ref{tab:ferm-carichi}. The threshold effects are not considered, because they are model dependent and we try to make a general analysis valid for the various models. In these theories there are no constrains on the value of $tan\beta$, so we use the high scale evolved data in the case of $tan\beta=10$.

\begin{table}[htbp]
\begin{small}
\begin{center}
\begin{tabular}{|l|r|}
\hline
Observables & Input data \\
\hline
\hline
$m_u [MeV]$ & $0.55 \pm 0.25$\\
\hline
$m_c [MeV]$ & $210 \pm 21$\\
\hline
$m_t$ [GeV]& $82.4^{+30.3}_{-14.8}$\\
\hline
$m_d [MeV]$ & $1.24 \pm 0.41$\\
\hline
$m_s [MeV]$ & $21.7 \pm 5.2$\\
\hline 
$m_b [GeV]$ & $1.06^{+0.14}_{-0.09}$\\
\hline
$m_e [MeV]$ & $0.3585 \pm 0.0003$\\
\hline
$m_\mu [MeV] $ & $75.672 \pm 0.058$\\
\hline
$m_\tau [GeV]$ & $1.2922 \pm 0.0013$\\
\hline
$V_{us}$ & $0.2243 \pm 0.0016$\\
\hline
$V_{cb}$ & $0.0351 \pm 0.0013$\\
\hline
$V_{ub}$ & $0.0032 \pm 0.0005$\\
\hline
$J\times 10^{-5}$ & $2.2 \pm 0.6$\\
\hline
\end{tabular}
\end{center}
\end{small}
\caption{GUT scale data for charged fermions for $tg\beta=10$ (ref.\cite{berto},\cite{parida})}
\label{tab:ferm-carichi}
\end{table}

For neutrino masses and PMNS mixings we use the results in Tab. \ref{tab:neutrino}. These values are obtained with a global fit considering also the recent results from T2K and MINOS. In the models we considered we never obtain degenerate neutrino mass spectrum, so the effects of the evolution from the low energy scale to the GUT scale can be considered negligible to a good approximation for these observables.

\begin{table}[htbp]
\begin{small}
\begin{center}
\begin{tabular}{|l|r|}
\hline
Observable & Input data \\
\hline
\hline
$\Delta m^2_{21}\times 10^{-5} [eV^2]$ & $7.59^{+ 0.20}_{-0.18}$ \\
\hline
$\Delta m^2_{31}\times 10^{-3} [eV^2]$ & $2.50^{+0.09}_{-0.16}$ \\
\hline
$sin^2\theta_{13}$& $0.013^{+0.007}_{-0.005}$ \\
\hline
$sin^2\theta_{12}$& $0.312^{+0.017}_{-0.015}$ \\
\hline
$sin^2\theta_{23}$& $0.52^{+0.06}_{-0.07}$ \\
\hline
\end{tabular}
\end{center}
\end{small}
\caption{Neutrino masses and mixing in normal hierarchy (ref. \cite{Schwetz:2011zk})}
\label{tab:neutrino}
\end{table}

\newpage

\section*{Appendix C}

Here we give some other details on the results of the numerical analysis. In particular for the three cases we analysed we give the best fit parameters and the values for the observables that we obtain. 

\begin{itemize}

\item {\it case A}:

\begin{eqnarray}
h&=&\left(
\begin{array}{ccc}
h_{11}&h_{12}\,e^{i \delta_{h_{12}}}&h_{13}\,e^{i \delta_{h_{13}}}\\
h_{12}\,e^{-i \delta_{h_{12}}}&h_{22}&h_{23}\,e^{i \delta_{h_{23}}}\\
h_{13}\,e^{-i \delta_{h_{13}}}&h_{23}\,e^{-i \delta_{h_{23}}}&h_{33}
\end{array}
\right) \\
h^\prime&=&\left(
\begin{array}{ccc}
h'_{11}&h'_{12}\,e^{i \delta_{h'_{12}}}&h'_{13}\,e^{i \delta_{h'_{13}}}\\
h'_{12}\,e^{-i \delta_{h'_{12}}}&h'_{22}&h'_{23}\,e^{i \delta_{h'_{23}}}\\
h'_{13}\,e^{-i \delta_{h'_{13}}}&h'_{23}\,e^{-i \delta_{h'_{23}}}&h'_{33}
\end{array}
\right) \\
f&=&\left(
\begin{array}{ccc}
f_2&f_1&f_1\\
f_1&f_2+f_0&f_1-f_0\\
f_1&f_1-f_0&f_2+f_0
\end{array}
\right)
\end{eqnarray}

\begin{table}[ht]
\begin{small}
\begin{center}
\subtable[]{
\begin{tabular}{|l|r|}
\hline
Observable & Best fit value \\
\hline
\hline
$m_u [MeV]$ & 0.550 \\
\hline
$m_c [MeV]$ & 210 \\
\hline
$m_t [GeV]$ & 81.9 \\
\hline
$m_d [MeV]$ & 1.24 \\
\hline
$m_s [MeV]$ & 21.7 \\
\hline 
$m_b [GeV]$ & 1.06 \\
\hline
$m_e [MeV]$ & 0.3585 \\
\hline
$m_\mu [MeV] $ & 75.67 \\
\hline
$m_\tau [GeV]$ & 1.292 \\
\hline
$V_{us}$ & 0.224 \\
\hline
$V_{cb}$ & 0.0351 \\
\hline
$V_{ub}$ & 0.00320 \\
\hline
$J\times 10^{-5}$ & 2.20 \\
\hline
$\Delta m^2_{21} \times 10^{-5} [eV^2]$ & 7.59 \\
\hline
$\Delta m^2_{32} \times 10^{-3} [eV^2]$ & 2.50 \\
\hline
$sin^2\theta_{13} $ & 0.0132 \\
\hline
$sin^2\theta_{12} $ & 0.312 \\
\hline
$sin^2\theta_{23} $ & 0.516 \\
\hline\hline
$\chi^2$ & 0.00500 \\
\hline
\end{tabular}
}\qquad\qquad
\subtable[]{
\begin{tabular}{|l|r|}
\hline
Parameter & Best fit value \\
\hline
\hline
$h_{11}v_u[GeV]$ & 1.40 \\
\hline
$h_{12}v_u[GeV]$ & -2.45 \\
\hline
$\delta_{h_{12}}$ & -1.39 \\
\hline
$h_{13}v_u[GeV]$ & 13.1 \\
\hline
$\delta_{h_{13}}$ & 0.232 \\
\hline
$h_{22}v_u[GeV]$ & 5.10 \\
\hline
$h_{23}v_u[GeV]$ & 15.0 \\
\hline
$\delta_{h_{23}}$ & 1.81 \\
\hline
$h_{33}v_u[GeV]$ & 79.1 \\
\hline
$h'_{11}v_u[GeV]$ & -6.42 \\
\hline
$h'_{12}v_u[GeV]$ & -6.22 \\
\hline
$\delta_{h'_{12}}$ & 0.901 \\
\hline
$h'_{13}v_u [GeV]$ & 2.62 \\
\hline
$\delta_{h'_{13}}$ & -0.652 \\
\hline
$h'_{22}v_u[GeV]$ & 4.55 \\
\hline
$h'_{23}v_u[GeV]$ & -31.4 \\
\hline
$\delta_{h'_{23}}$ & -1.06 \\
\hline
$h'_{33}v_u [GeV]$ & 27.3 \\
\hline
$f_0v_u[GeV]$ & -3.23 \\
\hline
$f_1v_u[GeV]$ & -0.155 \\
\hline
$f_2v_u[GeV]$ & 0.987 \\
\hline
$r_1/\tan{\beta}$ & -0.00418 \\
\hline
$c_e$ & -0.619 \\
\hline
$b_1$ & -1.12 \\
\hline
$b_{24}$ & 1.93 \\
\hline
$v_\nu/v_u \times 10^{-9}$ & 0.00947 \\
\hline
\end{tabular}
}
\end{center}
\end{small}
\caption{Fit result for the case A (26 parameters) described in Sect. 3}
\label{tab:fit-model1}
\end{table}

\newpage

\item {\it case B}:

\begin{eqnarray}
h^S&=&\left(
\begin{array}{ccc}
h_{11}&h_{12}&h_{13}\\
h_{12}&h_{22}&h_{23}\\
h_{13}&h_{23}&h_{33}
\end{array}
\right) \\
h^A&=&i\left(
\begin{array}{ccc}
0&\sigma_{12}&\sigma_{13}\\
\sigma_{12}&0&\sigma_{23}\\
\sigma_{13}&-\sigma_{23}&0
\end{array}
\right) \\
f&=&\left(
\begin{array}{ccc}
f_2&f_1&f_1\\
f_1&f_2+f_0&f_1-f_0\\
f_1&f_1-f_0&f_2+f_0
\end{array}
\right)
\end{eqnarray}

\begin{table}[ht]
\begin{small}
\begin{center}
\subtable[]{
\begin{tabular}{|l|r|}
\hline
Observable & Best fit value \\
\hline
\hline
$m_u [MeV]$ & 0.465 \\
\hline
$m_c [MeV]$ & 210 \\
\hline
$m_t [GeV]$ & 81.5 \\
\hline
$m_d [MeV]$ & 2.95 \\
\hline
$m_s [MeV]$ & 23.2 \\
\hline 
$m_b [GeV]$ & 1.09 \\
\hline
$m_e [MeV]$ & 0.3585 \\
\hline
$m_\mu [MeV] $ & 75.67 \\
\hline
$m_\tau [GeV]$ & 1.292 \\
\hline
$V_{us}$ & 0.224 \\
\hline
$V_{cb}$ & 0.0352 \\
\hline
$V_{ub}$ & 0.00321 \\
\hline
$J\times 10^{-5}$ & 2.15 \\
\hline
$\Delta m^2_{21} \times 10^{-5} [eV^2]$ & 7.59 \\
\hline
$\Delta m^2_{32} \times 10^{-3} [eV^2]$ & 2.50 \\
\hline
$sin^2\theta_{13} $ & 0.0118 \\
\hline
$sin^2\theta_{12} $ & 0.315 \\
\hline
$sin^2\theta_{23} $ & 0.516 \\
\hline\hline
$\chi^2$ & 5.64 \\
\hline
\end{tabular}
}\qquad\qquad
\subtable[]{
\begin{tabular}{|l|r|}
\hline
Parameter & Best fit value \\
\hline
\hline
$h_{11}v_u[GeV]$ & 0.785 \\
\hline
$h_{12}v_u[GeV]$ & 0.346 \\
\hline
$h_{13}v_u[GeV]$ & 7.83 \\
\hline
$h_{22}v_u[GeV]$ & 3.09 \\
\hline
$h_{23}v_u[GeV]$ & 4.35 \\
\hline
$h_{33}v_u[GeV]$ & 82.2 \\
\hline
$\sigma_{12}v_u[GeV]$ & -0.298 \\
\hline
$\sigma_{13}v_u [GeV]$ & 0.229 \\
\hline
$\sigma_{23}v_u[GeV]$ & 2.29 \\
\hline
$f_0v_u[GeV]$ & -0.935 \\
\hline
$f_1v_u[GeV]$ & 0.173 \\
\hline
$f_2v_u[GeV]$ & 0.0316 \\
\hline
$r_1/\tan{\beta}$ & 0.0132 \\
\hline
$c_e$ & 1.15 \\
\hline
$r_2$ & 2.50 \\
\hline
$v_\nu/v_u \times 10^{-9}$ & 0.0249 \\
\hline
\end{tabular}
}
\end{center}
\end{small}
\caption{Fit result for the case B (16 parameters) described in Sect. 3}
\label{tab:fit-model1}
\end{table}

\newpage

\item {\it case C}:

\begin{eqnarray}
h^S&=&\left(
\begin{array}{ccc}
h_{11}&h_{12}&h_{13}\\
h_{12}&h_{22}&h_{23}\\
h_{13}&h_{23}&h_{33}
\end{array}
\right) \\
h^A&=&i\left(
\begin{array}{ccc}
0&\sigma_{12}&\sigma_{13}\\
\sigma_{12}&0&\sigma_{23}\\
\sigma_{13}&-\sigma_{23}&0
\end{array}
\right) \\
f&=&\left(
\begin{array}{ccc}
f_2&f_1&f_1\\
f_1&f_2+f_0&f_1-f_0\\
f_1&f_1-f_0&f_2+f_0
\end{array}
\right)
\end{eqnarray}

\begin{table}[ht]
\begin{small}
\begin{center}
\subtable[]{
\begin{tabular}{|l|r|}
\hline
Observable & Best fit value \\
\hline
\hline
$m_u [MeV]$ & 0.550 \\
\hline
$m_c [MeV]$ & 210 \\
\hline
$m_t [GeV]$ & 82.2 \\
\hline
$m_d [MeV]$ & 1.24 \\
\hline
$m_s [MeV]$ & 21.6 \\
\hline 
$m_b [GeV]$ & 1.06 \\
\hline
$m_e [MeV]$ & 0.3585 \\
\hline
$m_\mu [MeV] $ & 75.67 \\
\hline
$m_\tau [GeV]$ & 1.292 \\
\hline
$V_{us}$ & 0.224 \\
\hline
$V_{cb}$ & 0.0351 \\
\hline
$V_{ub}$ & 0.00320 \\
\hline
$J\times 10^{-5}$ & 2.20 \\
\hline
$\Delta m^2_{21} \times 10^{-5} [eV^2]$ & 7.59 \\
\hline
$\Delta m^2_{32} \times 10^{-3} [eV^2]$ & 2.50 \\
\hline
$sin^2\theta_{13} $ & 0.0131 \\
\hline
$sin^2\theta_{12} $ & 0.312 \\
\hline
$sin^2\theta_{23} $ & 0.520 \\
\hline\hline
$\chi^2$ & 0.00149 \\
\hline
\end{tabular}
}\qquad\qquad
\subtable[]{
\begin{tabular}{|l|r|}
\hline
Parameter & Best fit value \\
\hline
\hline
$h_{11}v_u[GeV]$ & 0.584 \\
\hline
$h_{12}v_u[GeV]$ & -0.548 \\
\hline
$h_{13}v_u[GeV]$ & -5.49 \\
\hline
$h_{22}v_u[GeV]$ & 3.55 \\
\hline
$h_{23}v_u[GeV]$ & 3.99 \\
\hline
$h_{33}v_u[GeV]$ & 81.8 \\
\hline
$\sigma_{12}v_u[GeV]$ & -0.317 \\
\hline
$\sigma_{13}v_u [GeV]$ & 2.79 \\
\hline
$\sigma_{23}v_u[GeV]$ & -7.09 \\
\hline
$f_0v_u[GeV]$ & -0.999 \\
\hline
$f_1v_u[GeV]$ & -0.207 \\
\hline
$f_2v_u[GeV]$ & 0.0290 \\
\hline
$r_1/\tan{\beta}$ & 0.0129 \\
\hline
$r_2$ & 1.85 \\
\hline
$r_3$ & 1.37 \\
\hline
$c_e^S$ & 1.48 \\
\hline
$c_e^A$ & 1.17 \\
\hline
$v_\nu/v_u \times 10^{-9}$ & 0.0287 \\
\hline
\end{tabular}
}
\end{center}
\end{small}
\caption{Fit result for the case C (18 parameters) described in Sect. 3}
\label{tab:fit-model1}
\end{table}

\end{itemize}

\end{document}